\documentstyle[aps,prl,multicol,epsf]{revtex}

\begin{document}

\title{A Theory for Fluctuations in Stock Prices and Valuation of their Options}

\author{Gemunu H. Gunaratne,$^{1,2}$ and Joseph L. McCauley$^{1}$}

\address{$^{1}$ Department of Physics,
                University of Houston,
                Houston, TX 77204}
\address{$^{2}$ The Institute of Fundamental Studies,
                Kandy 20000, Sri Lanka}
\maketitle

\begin{abstract}
A new theory for pricing options of a stock is presented. It is based 
on the assumption that while successive variations in return are 
uncorrelated, the frequency with which a stock is traded depends on the 
value of the return. The solution to the Fokker-Planck equation 
is shown to be an asymmetric exponential distribution, similar to those
observed in intra-day currency markets. The ``volatility smile," 
used by traders to correct the Black-Scholes pricing is shown to 
provide an alternative mechanism to implement the new options pricing 
formulae derived from our theory.
\end{abstract}
\nobreak
\begin{multicols}{2}

Although options contracts have been in use as far back as the reign 
of Hammurabi in Babylon~\cite{dunb}, the first useful theoretical analysis for their valuation 
was presented relatively recently~\cite{blaAsch,coxArub}.
In the Black-Scholes theory, variation in the price $S(t)$ of a stock
are investigated using the ``return" $x(t) = log [S(t)/S_0]$~\cite{osbo},
where $S_0$ is a ``consensus" value of the stock at the time ($t=0$) that
the option is purchased~\cite{consen}. $S_0$ is typically set to be the price $S(t=0)$. 
The expected growth rate $b$ for $S(t)$ satisfying 
\begin{equation}
   \left < S(t)\right> = S_0 e^{bt},
\label{growth1}
\end{equation}
may differ from the interest rate $r$ on funds borrowed to purchase the option. 
It is further assumed that successive random fluctuation in returns 
are independent and identically distributed, and hence 
(by central limit theorem) that $x(t)$ lies on a normal distribution for 
sufficiently large $t$~\cite{chan}.
A European call (i.e., an option to purchase a stock
at a ``strike" price $K$), is valued by its expected profit at expiration; 
i.e., $C_{BS}(S,K,t) = exp(-rt)\int_{K}^{\infty} dS (S-K) f_{LN}(S,t)$~\cite{coxArub}.
In the Black-Scholes theory, the distribution of stock prices $f_{LN}(S,t)$ 
is log-normal and it can be shown that~\cite{coxArub} 
\begin{equation}
C_{BS}(K,S_0,t) = S_0 e^{(b-r)t} {\cal N}(d_+) - K e^{-rt} {\cal N}(d_-).
\label{BS_call}
\end{equation}
Here ${\cal N}(x)$ denotes the cumulative normal distribution, 
\begin{equation}
d_{\pm} = \frac{bt+log(S_0/K)}{\sigma \sqrt{t}} \pm \frac{\sigma \sqrt{t}}{2}, 
\label{defnd}
\end{equation}
and $\sigma$ is referred to as the volatility of the return.
Similarly, a European put (i.e., an option to sell a stock at a price $K$), 
is valued using
$P_{BS}(S,K,t) = exp(-rt) \int_{0}^{K} dS (K-S) f_{LN} (S,t)$, and is given by 
\begin{equation}
P_{BS}(K,S_0,t) = K e^{-rt} {\cal N}(-d_-) - S_0 e^{(b-r) t} {\cal N}(-d_+).
\label{BS_put}
\end{equation}

\begin{figure}
\epsfxsize=2.0in
\hskip 0.4truein
\epsffile{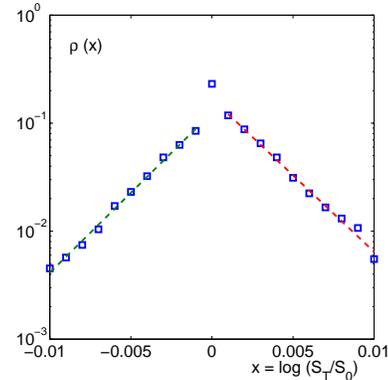}
\caption{The relative frequency of intra-day returns (delay of 4 hours) for 
U.S. bonds over a period of 600 consecutive trading days beginning in 
March, 1985. For returns shown, it is approximately an asymmetric 
exponential distribution.} 
\label{bonds}
\end{figure}

Recent investigations of bond and foreign exchange markets have clearly shown 
that the distribution of returns (for a fixed delay) 
deviates significantly from a normal distribution, 
especially far from the mean~\cite{manAsta,ghaAbre,friApei,arnAmuz,dacAgen,balAdac}. 
In fact, to a very good approximation, intra-day currency fluctuations lie on an 
an asymmetric exponential distribution, see Figure~\ref{bonds}~\cite{macAgun}.
It is also known that currency traders do not assign values for options 
according to Eqns.~(\ref{BS_call}) and (\ref{BS_put}). The most common correction is
the use of a heuristically deduced ``volatility smile"; i.e., a strike price 
dependent expression for $\sigma$ (which contradicts the basis of the 
Black-Scholes theory). In this Letter, we argue that the
use of a volatility smile can account for changes in the value of options 
due to deviation of the distribution of returns from normality.
This conclusion is reached via a new theory
that is constructed on the assumption that {\it although successive events 
are uncorrelated, the step size of the corresponding random walk depends 
on the values of the return $x(t)$ and time $t$.}

Analyses of financial markets have provided evidence for the assertion 
that successive variations of the return are uncorrelated~\cite{dacAgen,arnAbou}. 
It implies that the distribution function $W(x,t)$  for returns $x(t)$ satisfies a 
Fokker-Planck equation~\cite{fokApla,chan}
\begin{equation}
\frac{\partial W}{\partial t} = -B \frac{\partial W}{\partial x} + \frac{1}{2} 
  \frac {\partial^2} {\partial x^2} (DW),
\label{FPeqn}
\end{equation}
$D\equiv D(x,t)$ being the diffusion coefficient. 
The drift rate $B$ is assumed to be constant for 
intra-day fluctuations~\cite{other}. Following conclusions from studies of financial 
markets~\cite{manAsta,dacAgen}, we limit our considerations to solutions 
that have a scaling form,
\begin{equation}
 W(x,t) = \frac{1}{t^{\eta}} F(u).
\label{scaling}
\end{equation}
Here $u=x/t^{\eta}$, and $\eta$ is referred to as the ``drift exponent."
Notice that the pre-factor $t^{-\eta}$ is introduced in order that $W(x,t)$ is
normalized for all $t$.

Consider first the case $B=0$. Financial markets exhibit the following
behavior; a stock whose price $S(t)$ deviates significantly (up or down)
from $S_0$ is traded at a higher frequency; i.e., the local diffusion rate $D(x,t)$ 
is enhanced. This effect is further amplified if the variation occurs during a shorter
time interval. These observations are quantified in an 
assumption that $D(x,t)$ is a bi-linear function of $u$,
\begin{equation}
  D(x,t) = \frac{1}{\gamma^2} (1-\gamma u) \bar\Theta(u) 
         + \frac{1}{\nu^2} (1+\nu u) \Theta(u).
\label{diff}
\end{equation}
Here $\Theta(u)$ denotes the Heaviside $\Theta$-function and $\bar\Theta(u) = 1-\Theta(u)$. 
As will become apparent momentarily, for $W(x,t)$ to be asymmetric,
it is necessary for the parameters $\gamma$ and $\nu$ to be different.
Notice that for larger values of $u$, nonlinear corrections to $D(x,t)$ 
may be required.

These assumption on the absence of correlations between successive movements
of the return, and the scaling hypotheses (\ref{scaling}) and (\ref{diff})
provide a unique value for $\eta$. This can be seen from 
Eqn.~(\ref{FPeqn}), which simplifies to
\begin{equation}
  -\frac{\eta}{t^{\eta+1}} F(u) - \frac{\eta}{t^{\eta+1}} u F'(u)
      = \frac{1}{2} \frac{1}{t^{3\eta}} (DF)^{\prime\prime}(u).
\label{wander}
\end{equation}
Consequently $\eta=1/2$, in agreement with conclusions from the
analysis presented in Ref.~\cite{manAsta} which
show that the standard deviation of $W(x,t)$ for the S\&P500 time series increases 
approximately as $\sigma (t) \sim t^{0.53}$.

Next, substituting the diffusion coefficient (\ref{diff}) in 
Eqn.~(\ref{wander}) for $x> 0$ gives,
\begin{equation}
\left(\frac{1+\nu u}{\nu^2}\right) F^{\prime\prime}(u) 
      + \left( u +\frac{2}{\nu} \right) F^{\prime} (u) + F(u) = 0.
\label{solnu}
\end{equation}
Writing $F(u) = \sum_{n=0} a_n u^{\beta+n}$~\cite{benAorz}, it is 
found that $F(u) \sim exp(-\nu u)$ is one of the solutions~\cite{commnt}. 
Combining it with the corresponding analysis for $x<0$ provides 
the solution
\begin{equation}
  W(x,t) = \frac{A}{\sqrt{t}} e^{\gamma u} \bar\Theta(u) + 
           \frac{B}{\sqrt{t}} e^{-\nu u} \Theta(u)  
\label{distrn}
\end{equation} 
to the Fokker-Planck equation.
Two conditions are required to evaluate the coefficients $A$ and $B$.
Normalizing $W(x,t)$ gives 
\begin{equation}
   \frac{A}{\gamma} + \frac{B}{\nu} = 1.
\label{prefac1}
\end{equation}
The second condition imposed is the continuity of the probability current
$j(x,t) = -(DW)_x$, without which probability would accumulate at the 
origin. Using $\Theta^2(u)= \Theta(u)$, $\Theta_u(u)=\delta(u)$, 
$\bar\Theta_u(u) = -\delta(u)$ and $f(u) \delta(u) = f(0) \delta(u)$,
\begin{eqnarray}
j(x,t) = &-& \frac{\gamma^2 u e^{\gamma u}}{(\gamma+\nu)t} \bar\Theta(u) 
        - \frac{\nu^2 u e^{-\nu u}} {(\gamma + \nu) t} \Theta (u) \nonumber \\
        &+& \frac{1}{t} \left(\frac{B}{\nu^2} - \frac{A}{\gamma^2}\right) \delta(u).
\label{deriv}
\end{eqnarray}
Hence, continuity of $j(x,t)$ at $u=0$ implies that
\begin{equation}
   \frac{A}{\gamma^2} = \frac{B}{\nu^2}.
\label{prefac2}
\end{equation}
Thus  $A=\gamma^2 / (\gamma + \nu)$ and $B = \nu^2 / (\gamma+\nu)$. Notice that 
$W(x,t)$ contains a discontinuity at the origin of $(\gamma-\nu)/\sqrt{t}$.
Notice also that the mean value $\left< x(t)\right> = 0$ for all $t$
and that the variance is $2t/\gamma \nu$; in particular,
$W(x,t=0) = \delta(x)$, as required by the initial condition.

We make two further observations. If $D(x,t)$ is translated by 
$\Delta$, then the corresponding solution to the Fokker-Planck equation
is $W(x-\Delta, t)$; however, the requirement $\left< x(t=0) \right> = 0$ 
forces $\Delta=0$. Second, the solution for the general case 
(i.e.,  $B \neq 0$) is obtained by Galilean transformations of $D(x,t)$ 
and $W(x,t)$. This can be achieved by replacing $u$ by
$(x-Bt)/\sqrt{t}$.  For this case $\left< x(t) \right> = Bt$.

These conclusions were tested by investigating the following two types of
random walks (for the case $B=0$). In the first, fixed-step-size walk
(i.e., each step has a unit magnitude),  
the time interval for a step from a location $x(t)$ is chosen 
to be $\gamma^2 /(1-\gamma u)$ if $u<0$ and to be $\nu^2 / (1+\nu u)$ if $u\ge 0$.
In the second, fixed-step-time walk (i.e., each step takes a 
unit time) the size of a step is chosen to be 
$\sqrt{(1-\gamma u)} / \gamma$ or $\sqrt{(1+\nu u)} / \nu$ depending on 
whether $u<0$ or $u \ge 0$. In each case the motion begins at the origin, 
and the direction of each step is chosen 
randomly with equal probability. The effective diffusion coefficient for 
these random walks is given by Eqn.~(\ref{diff})~\cite{chan}.

\begin{figure}
\epsfxsize=2.0in
\hskip 0.4truein
\epsffile{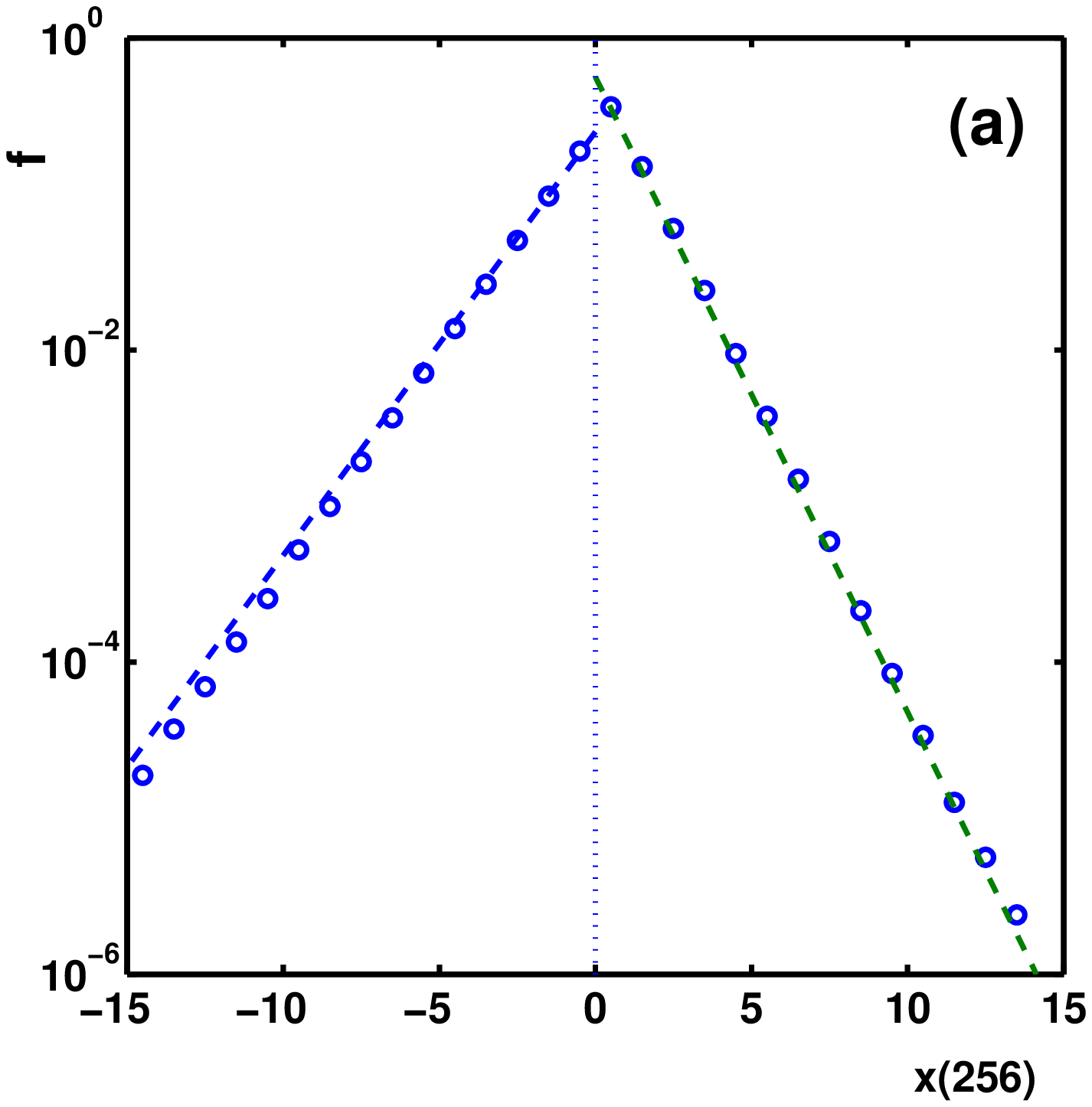}
\vskip 0.1truein
\epsfxsize=2.0in
\hskip 0.4truein
\epsffile{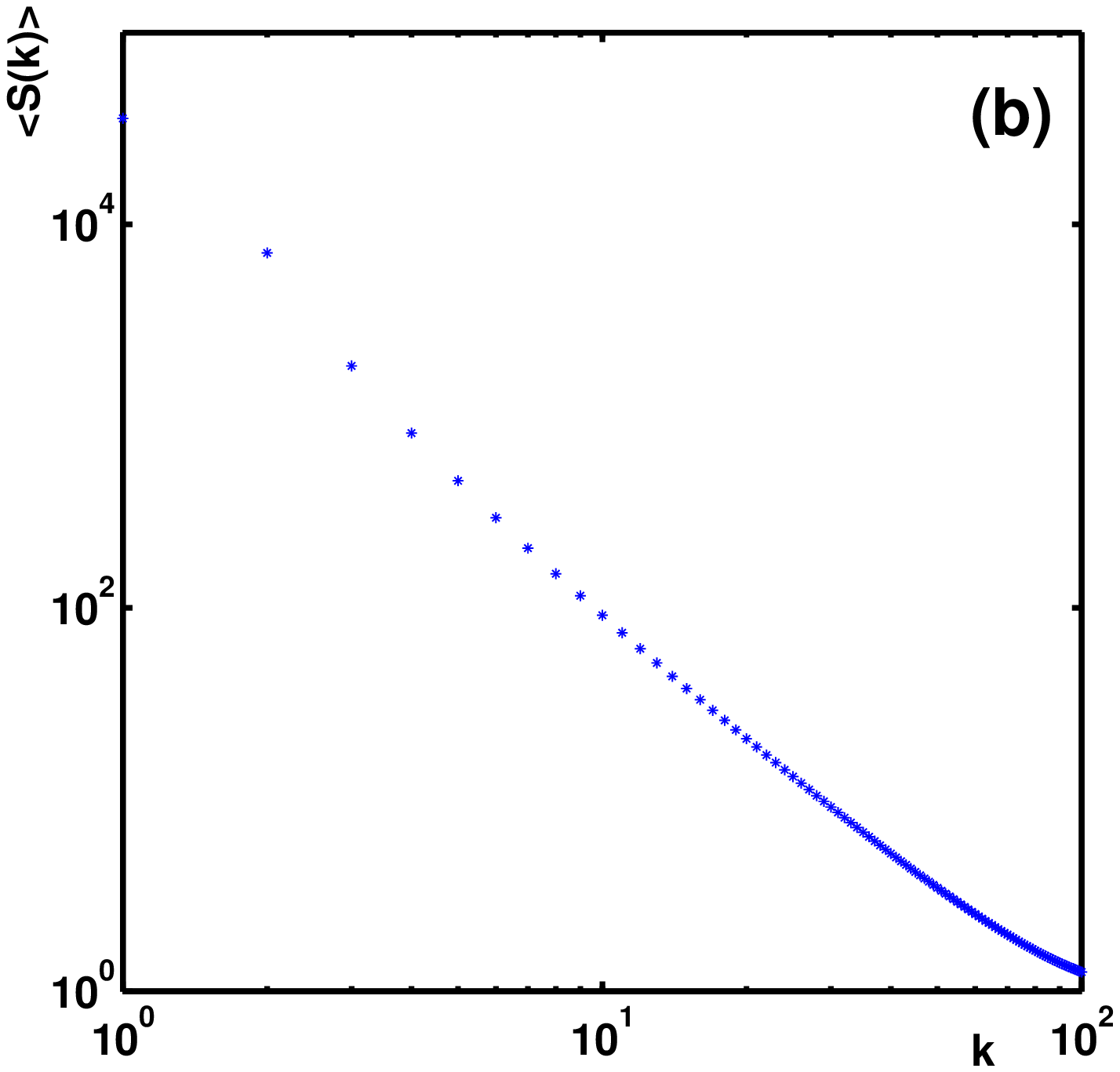}
\caption{(a) The relative frequency of final locations for 5 million fixed-step-time walks of
256 steps. In this example $\gamma=10.0$ and $\nu=15.0$. The dashed lines show the
distribution (\ref{distrn}).
(b) The mean $\left<S(k)\right>$ of the 
power spectrum for these random walks exhibits a $k^{-2}$ decay, 
similar to that observed in analyses of financial markets.}
\label{hist}
\end{figure}

Figure~\ref{hist}(a) shows the histogram of positions for 5 million
fixed-step-time walks of 256 steps, which is clearly consistent
with $W(x,t)$ given by Eqn.~(\ref{distrn}). 
The mean value of the power spectra for these random walks, shown in 
Figure~\ref{hist}(b), exhibits a 
$k^{-2}$ decay; a similar behavior has been reported in an analysis
of financial markets~\cite{manAsta,spect}. 

One final point needs to be clarified prior to deriving formulae for valuation of options.
Condition (\ref{growth1}) implies that
\begin{equation}
   G \equiv  \frac{\gamma\nu - (\gamma-\nu)\sqrt{t}} {(\gamma+\sqrt{t})(\nu-\sqrt{t})} 
        = e^{(b-B)t}.
\label{delta}
\end{equation}
Since this is not an identity, the equality cannot be valid
for all $t$; hence a growth rate given by (\ref{growth1}) is not 
consistent with the new theory. We propose to replace it by
\begin{equation}
\left<x(t)\right> = \left< \log \frac {S(t)}{S_0}\right> = bt.
\label{growth2}
\end{equation}
This last condition imposed on our theory implies that the parameter $B$ in the
Fokker-Planck equation is $b$.  

Now we are in a position to re-derive the formulae for pricing of options.
Consider first, a European call to be exercised at time $t$. Pricing the
option by its expected profit, it is found that
\begin{eqnarray}
C e^{rt} &=& e^{bt} G - K + K \zeta \frac{\gamma}{\gamma+\sqrt{t}}  
              z^{\gamma/\sqrt{t}},  \ \ \ z<1 \nonumber \\
&=& K \zeta \frac{\nu}{\nu-\sqrt{t}} z^{-\nu/\sqrt{t}}, 
                      \ \ \ \ \ \ \ \ \ \ \ \ \ \ \ \ \  z\ge 1 
\label{callpr}
\end{eqnarray}
where $z= (K/S_0)exp(-bt)$ and $\zeta = \sqrt{t}/(\gamma+\nu)$.
 
A European put can be valued using the definition 
$P(S_0, K, t) = e^{-rt} \int_{0}^{K} dS (K-S) f(S,t)$. 
When the return is distributed according to Eqn.~(\ref{distrn}), 
\begin{eqnarray}
P e^{rt} &=& K-e^{bt} G + K \zeta \frac{\nu}{\nu-\sqrt{t}} z^{-\nu/\sqrt{t}},
                                  \ z > 1 \nonumber \\
&=& K \zeta \frac{\gamma}{\gamma+\sqrt{t}} z^{\gamma/\sqrt{t}}. 
                      \ \ \ \ \ \ \ \ \ \ \ \ \ \ \ \ \ \ \ z \le 1
\label{putpr}
\end{eqnarray}

\begin{figure}
\epsfxsize=2.0in
\hskip 0.40truein
\epsffile{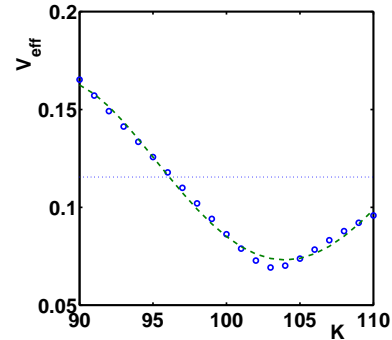} 
\caption{The effective volatility evaluated using Eqn.~(\ref{voldef}) for
several values of the strike price $K$, in the neighborhood of the underlying stock 
price of 100 units.  $\gamma$ and $\nu$ are 10.0 and 15.0 respectively, and
the time to expiration of the option is half a year. The dashed line shows
a quartic fit to the data, and the dotted line the standard deviation
for the distribution.}
\label{volsmile}
\end{figure}

Eqns.~(\ref{callpr}) and (\ref{putpr}) can be used to justify the use of 
volatility smile curves to correct Black-Scholes pricing formulae.
Suppose the ``true" distribution of stock returns is (\ref{distrn}),
which decays slower than a Gaussian. Larger deviations
of $x(t)$ will occur more frequently than predicted by the Gaussian 
distribution under this scenario (and will be noted by an observant trader).
As a result, these options will be valued higher than the pricing 
given by the Black-Scholes theory. For a given strike price,
this enhancement can be accounted for by increasing the ``effective"
volatility (used in the Black-Scholes formulae). As an example, 
consider a distribution $W(x,t)$ with $\gamma=15.0$ and $\nu=10.0$.
(The annualized volatility for this distribution is 11.5\%.) The ``correct" 
valuation for options is assumed to be given by Eqns.~(\ref{callpr}) and ~(\ref{putpr}).
We assign an effective volatility $V_{eff}$ at a strike price $K$ by
equating the sum of prices of a call and a put estimated from the 
Black-Scholes theory (i.e., Eqns.~(\ref{BS_call}) and (\ref{BS_put})) with 
the corresponding sum for the new theory; i.e., $V_{eff}(K)$ is chosen
so that 
\begin{equation}
P_{BS}(S_0, K, t) + C_{BS}(S_0, K, t) = P(S_0, K, t) + B(S_0, K, t).
\label{voldef}
\end{equation} 
The function $V_{eff}(K)$ (for $S_0= 100$, $t=0.5$ years, $b=10\%$, and
$r$ arbitrary), shown in Figure~\ref{volsmile}, resembles volatility
smile curves used in financial trading.

We conclude by elaborating on a couple of issues raised by our work.
The consensus value of a stock depends on many factors such as its
historical performance and the market's expectations for its future
prospects. In the Black-Scholes theory, it is used to define the returns.
Typically, $S_0$ is chosen to be the price of the stock at $t=0$. 
If a different value is used for $S_0$ (with a suitable modification of
Eqn.~(\ref{growth1})), the only effect is a uniform shift
of $W(x,t)$; the pricing formulae remain unchanged. (Another way to 
state this is that the Langevin equation~\cite{chan} is independent
of $S_0$.)  In contrast, $S_0$ plays a unique role in our theory. 
It is the value of the stock at which the diffusion coefficient 
reaches a minimum. Any deviation of the stock price from this value
is reflected in an increase in the magnitude of its fluctuations. 
(The Langevin equation depends on $S_0$ through the diffusion 
coefficient $D(x,t)$.)

\begin{figure}
\epsfxsize=2.0in
\hskip 0.40truein
\epsffile{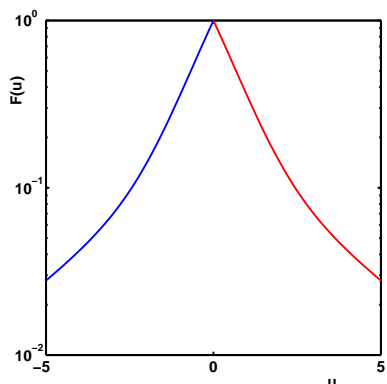} 
\caption{A solution $F(u)$ of the Fokker-Planck equation with a diffusion 
coefficient that includes a quadratic term in $u$. For this example, $\gamma = \nu=1.0$,
and $\epsilon=-0.2$.}
\label{fattail}
\end{figure}

The second issue concerns possible higher order corrections to the
bilinear form (\ref{diff}) for the diffusion coefficient. For example, if
a quadratic term $\gamma^{-2} \epsilon u^2$ is added to $D(x,t)$ 
for $x>0$. Then, the ordinary differential equation satisfied by 
$F(u)$ changes to
\begin{eqnarray}
(1 + \nu u + \epsilon u^2) F^{\prime\prime}(u) &+& (2\nu + 4\epsilon u + \nu^2 u) F^{\prime} (u) 
                                                                                \nonumber\\
  &+& (2\epsilon + \nu^2) F(u) = 0.
\label{perturb}
\end{eqnarray}
The solution for the symmetric case ($\gamma = \nu$) to this equation that 
satisfies the initial condition $F^{\prime} (0+) = -\nu F(0)$ is 
shown in Figure~\ref{fattail}. $F(u)$ can be shown to decay as a power
law for large $u$. Observations of such ``fat-tails"  have been
made on distributions of inter-day returns~\cite{dacAgen,manAsta}.
In spite of such possible corrections to the distribution $W(x,t)$, we believe 
that the theory presented here is still relevant in determining valuation of 
options. This is because, current frequency of trades (and not long time 
behavior of markets) is the information immediately available to traders,
and on which their decisions are based. 
Hence, one may expect that distributions of intra-day variations in the 
return will determine the valuation of options.

The authors would like to thank K. E. Bassler, M. Dacorogna, G. F. Reiter and
H. Thomas for discussions and their suggestions. GHG would also 
like to thank his colleges at TradeLink Corporation during 1989-1990.
This research is partially funded by the National Science Foundation 
and the Office of Naval Research (GHG).

\end{multicols}

\begin{thebibliography}{99}
\bibitem{consen} This is the price assigned by the market on the basis of factors
  such as the historical performance and future expectations for the growth of a stock.
\bibitem{commnt} The pre-factors in diffusion coefficient (\ref{diff}) was chosen 
  in order to get an exponential solution for $W(x,t)$.
  The second solution to Eqn.~(\ref{solnu}) is given by 
  $F(u) = \sum_{n=0} (-u)^{n}c_n / (n+1)!$ where $c_{n+1}=c_n + n!$.
\bibitem{spect} This form is observed for small values of $k$. Generally, the spectrum
  is proportional to $\sin^{-2}(k/2)$.
\end{thebibliography}
\end{document}